\documentclass[twocolumn,showpacs,preprintnumbers,amsmath,amssymb]{revtex4}

\usepackage{graphicx}
\usepackage{dcolumn}
\usepackage{bm}

\begin{document}

\preprint{APS/123-QED}

\title{Universality of solar wind turbulent spectrum from MHD to electron scales}

\author{O. Alexandrova}
\email{alex@geo.uni-koeln.de}
 \altaffiliation{Institute of Geophysics and Meteorology, University of Cologne, Albertus-Magnus-Platz 1,  50923, Cologne, Germany.}
\author{J. Saur}
\affiliation{Institute of Geophysics and Meteorology, University of Cologne, Albertus-Magnus-Platz 1,  50923, Cologne, Germany.}
\author{C. Lacombe}
\affiliation{LESIA, Observatoire de Paris, CNRS, UPMC, Universit\'e Paris Diderot, 5 place J.~Janssen, 92190 Meudon, France.}
\author{A. Mangeney}
\affiliation{LESIA, Observatoire de Paris, CNRS, UPMC, Universit\'e Paris Diderot, 5 place J.~Janssen, 92190 Meudon, France.}
\author{J. Mitchell}
\affiliation{Blackett Laboratory, Imperial College London, London SW7 2AZ, UK.}
\author{S. J. Schwartz}
\affiliation{Blackett Laboratory, Imperial College London, London SW7 2AZ, UK.}
\author{P. Robert}
\affiliation{LPP, 10--12 avenue de l'Europe 78140 Velizy France.}
\date{\today}

\begin{abstract}
In order to investigate the universality of magnetic turbulence in space plasmas we analyze seven time periods in the free solar wind of different origin, slow or fast, and under different plasma conditions. The orientation of magnetic field to the flow velocity was always quasi-perpendicular. Unique combination of three instruments on Cluster spacecraft which operate in different frequency ranges give us the possibility to resolve spectra up to 300~Hz. We show that spectra measured under different plasma conditions have a similar shape. Such a quasi-universal spectrum consists of three parts: two power laws and an exponential domain. At MHD scales, Kolmogorov's law $\sim k^{-5/3}$ is found. At scales smaller than the ion characteristic scales, a $k^{-2.8}$ law is observed. At scales $k\rho_e\sim (0.1-1)$, where $\rho_e$ is  the electron gyroradius, the magnetic spectrum follows an exponential law $\exp(-k^{1/2})$, indicating the onset of dissipation. This is the first observation of an exponential magnetic spectrum in space plasmas. We show that among several spatial kinetic plasma scales,  the electron Larmor radius plays the role of a dissipation scale in space plasma turbulence.
\end{abstract}

\maketitle

Space plasmas are usually in a turbulent state, and  the solar wind is one of the closest laboratories of space plasma turbulence, where in-situ measurements are possible thanks to a number of space missions  \cite{noi}. These measurements obtain time series which  provide  access to frequency spectra or to spectra of wave vectors along the flow.  It is  well established that at MHD scales (below $\sim 0.3$~Hz, at $1$~AU) the solar wind turbulent spectrum of magnetic  
fluctuations follows the Kolmogorov's spectrum $\sim f^{-5/3}$.  However, the characteristics of turbulence in the vicinity of the kinetic plasma scales (such as the inertial lengths $\lambda_{i,e} = c/\omega_{pi,e}$, $c$ being the speed of light and $\omega_{pi,e}$ the plasma frequencies of ions and electrons, respectively, the Larmor radii $\rho_{i,e}$ and the cyclotron frequencies $\omega_{ci,e} = eB/m_{i,e}$) are not well known experimentally and are a matter of debate. It was shown that at ion scales the turbulent spectrum has a break, and 
steepens to $\sim f^{-s}$, with a spectral index $s$ that is
clearly non-universal, taking on values in the range $-4$ to $-2$ \citep{Leamon1998,Smith2006}. These indices  were obtained  from data that enabled a rather restricted range of scales above the break to be investigated, up to $\sim 3$~Hz.  It is not known whether such indices persist at higher frequencies. At electron scales, the observations are difficult and our knowledge is very poor.  \citet{Denskat1983}  using Helios data obtained high resolution magnetic spectra at 2 distances from the Sun:  up to $50$~Hz at $1$~AU, and up to $470$~Hz at $0.3$~AU. However, in both cases, the electron characteristic scales were not reached. 
It was only with Cluster observations that these electron scales were reached. For the solar wind downstream of the Earth's bow-shock, it was shown that the turbulence spectrum changes its shape around $k\lambda_e\simeq k\rho_e\sim 1$ \cite{alexandr08angeo}. This result was recently confirmed in the upstream solar wind magnetically connected to the bow-shock \cite{Fouad09}. However, in both studies  the plasma $\beta$ (the ratio between plasma and magnetic pressures) was $\sim 1$ and so it was not possible to separate the roles of  $\rho_e$ and $\lambda_e$. 

Measurements of solar wind turbulent spectra in the vicinity of ion and electron plasma scales may clarify our understanding of the processes of dissipation (or dispersion) of turbulent energy in collisionless plasmas. A number of processes may be considered at these scales: cyclotron damping at $f_{ci}$  and $f_{ce}$  of Alfv\'en and whistler waves, respectively \cite{Gary-book}; scattering of oblique whistler waves at   $f_{ci}<f<f_{ce}$  \cite{Gogoberidze2005};  linear dissipation of kinetic  Alfv\'en waves at $1/\rho_i<k<1/\rho_e$ \cite{Bale2005,Schekochihin2009}.   

In this paper we use the Cluster spacecraft \cite{Escoubet1997} data to analyze the free solar wind of different origin, fast and slow, and under different plasma conditions. While Sahraoui et al. \cite{Fouad09} use the FluxGate Magnetometer (FGM) \citep{Balogh2001} and STAFF-Search Coil (SC) \citep{Cornilleau2003} at the burst mode, which allow them in principle to investigate turbulence spectra up to $180$~Hz, we complete these instruments with the STAFF-Spectra Analyzer (SA),  enabling us  to increase considerably the upper frequency limit, up to $4$~kHz. However, as was shown in \cite{Fouad09}, above $100$~Hz the instrument noise becomes a significant issue which we take into account in our analysis.  

As suggested in \citep{Cornilleau2003}, we use measurements in the magnetospheric lobe (precisely, the data on 5 April 2001,  $06$:$00$-$07$:$00$~UT) as  the noise level of the instrument. The final spectra were obtained by  substracting the lobe spectrum from the solar wind spectra. A similar procedure has been applied by \citet{Lin1998} for the Ulysses spacecraft data.  The maximal frequency in our analysis is defined as  the highest frequency where the measured spectrum is higher than twice the lobe spectrum, before subtraction. 

We select seven time intervals of 42 minutes when Cluster was at apogee (19 Earth radii) and spent one hour or more in the free solar wind: the electric field data at the electron plasma frequency show no evidence of magnetic connection to the bow shock [P.~Canu, private communication, 2009]. In Table~I, the dates of the intervals are shown as ymmdd, and their starting times are denoted by $t_i$. Average plasma parameters for the selected intervals are given also in this table.  Magnetic field measurements were obtained  from Cluster~1. Ion moments (density $N$, velocity $V$ and perpendicular temperature $T_{\perp i}$) are measured by the CIS/HIA experiment \cite{CIS2001} on Cluster~1.  The ion parallel temperatures are not properly determined in the solar wind by the CIS instrument  [I. Dandouras, private communication, 2009].  Electrons are measured by the PEACE instrument \cite{Johnstone1997}, mostly on Cluster~2.  One can see from Table~I that the mean field/flow angle, $\Theta_{BV}$, is always larger than $60^{\circ}$. Other plasma parameters are rather variable:  $V$ varies from  $\sim 360$~km/s to $670$~km/s, the total perpendicular plasma beta, $\beta_{i\perp}+\beta_{e\perp}=2\mu_0 nk(T_{i\perp}+T_{e\perp})/B^2$, varies between 0.7 and 3.3, the Alfv\'en speed  $V_a\in [30,130]$~km/s. $V_{th i,e} = \sqrt{kT_{\perp i,e}/m_{i,e}}$ are the ion and electron perpendicular thermal speeds, respectively; $\rho_{i,e}=V_{th i,e}/\omega_{ci,e}$ are the corresponding Larmor radii.  During these seven intervals we never observe quasi-parallel whistler waves, characterized by a quasi-circular right-hand polarization, which can be captured by STAFF-SA instrument. The two intervals 3 and 5 display the most intense spectra and are observed in the fast solar wind, a few hours downstream of an interplanetary shock.

\begin{table}
\caption{Solar wind parameters for selected time periods.\label{tbl-1}}
\begin{ruledtabular}
\begin{tabular}{lrrrrrrr}
Nb & 1 & 2 & 3 & 4 & 5 & 6 & 7 \\
ymmdd & 10405 & 20219 & 30218 & 31231 & 40122 & 40127 & 50112 \\ 
$t_i$(UT) & 22:36 &01:48 &00:18 & 10:48 & 05:03 & 00:36 & 02:00 \\
\hline
$B$(nT) & 7.3 &   7.0&  15.5 &   10.9 &  15.5 &  9.5 & 13.6 \\
$N$(cm$^{-3}$) &       3 &      29 &   7 &  22 &  20 &  8 &  33 \\
$T_{\perp i}$(eV)&  17 &  7 &  40 &  10&  61 &  10 &  14 \\
$T_{\perp e}$(eV)& 36 &  7 &  18 & 16 &  28&  21 & 16 \\
$V$(km/s)    &  540 & 365 & 670 &  430 & 635 & 430 & 440 \\
$\Theta_{BV}$($^{\circ}$) & 85 &  65  &   80 &     75 &     85  &    80 &      85 \\
\tableline
$\beta_{\perp i}$&  0.4 &  1.7 &  0.5&   0.8 &  2.0&  0.4 &  1.0 \\
$\beta_{\perp e}$&  0.8&  1.6& 0.2&  1.2&  0.9&  0.8&  1.2\\
\tableline
$f_{ci}$(Hz)   &  0.11  & 0.11   &  0.24   &  0.17  &  0.24  &  0.15  &  0.21 \\
$f_{ce}$(Hz)  & 205 & 195 &  435 & 305 & 435 & 265 & 380 \\
\tableline
$\lambda_i$(km)& 130 & 40 & 85 & 50 & 50 & 80 & 40 \\
$\lambda_e$(km)&       3  &     1  &    2   &   1  &    1 &     2  &   1 \\
$\rho_i$(km)& 60 & 40 & 40 & 30 & 50 & 35 & 30 \\
 $\rho_e$(km)& 2 & 1 & 0.5 & 1 & 1 & 1 & 0.5 \\
\tableline
$V_{a}$(km/s) &      95  &    30 &     130 &     50  &    75 &      70 &   50 \\
$V_{thi}$(km/s) &       40   &   25    &  60    &  30  &   75   &   30  &   35 \\
$V_{the}10^{3}$(km/s) & 2.5 &1.1 &1.8  & 1.7  & 2.2  & 1.9 & 1.7\\
\end{tabular}
\end{ruledtabular}
\end{table}

\begin{figure}
\includegraphics[width=6.5cm]{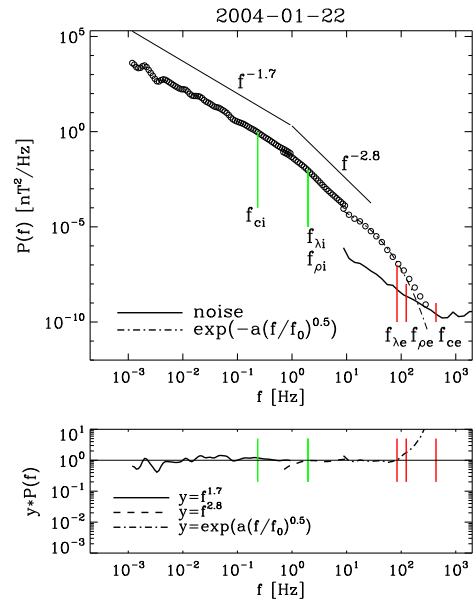}
\caption{\label{fig:spec} Top: Magnetic power spectral density for interval 5, measured by Cluster/FGM (up to $1$~Hz), STAFF--SC (up to $10$~Hz) and STAFF--SA ($f\geq 8$~Hz) instruments in the solar wind. Vertical bars  indicate plasma kinetic scales, where $f_{\lambda_{i,e}}$ correspond to the Doppler-shifted $\lambda_{i,e}$  and $f_{\rho_{i,e}}$  to $\rho_{i,e}$. Power-laws $f^{-1.7}$ and $f^{-2.8}$ are shown. Dashed-dotted line indicates exponential fit $\sim \exp (-a(f/f_0)^{0.5})$, with $f_0=f_{\rho_{e}}$ and the constant $a\simeq 9$.  Bottom: Compensated spectrum by $f^{1.7}$ at low frequencies, $f^{2.8}$ for middle part, and by the exponential for high frequency part. }
\end{figure}

Figure~\ref{fig:spec}(top) shows the magnetic spectrum $P(f)$ for interval 5. It is calculated using the Morlet wavelet transform, as was done in \cite{alexandr08apj}. One can clearly recognize here two power-laws and an exponential ranges: At low frequencies, the spectrum is $\sim f^{-1.7}$ consistent with Kolmogorov's law. Between $f_{ci}$ and $f_{\lambda_i} \simeq f_{\rho_i}$  (where $f_{\lambda_{i}}=V/2\pi \lambda_{i}$ and $f_{\rho_i}=V/2\pi \rho_{i}$), the first break appears. At higher frequencies, the spectrum follows an $\sim f^{-2.8}$ law. However, at $10\leq f \leq 100$~Hz, 
the spectrum is no longer a power-law, but follows approximatively an exponential function $\exp (-a(f/f_0)^{0.5})$. At higher frequencies, $f >f_{\lambda_e}$, the spectrum is too close to the noise level (see the black solid line) to draw any firm conclusions.

To demonstrate the above scaling laws, Figure~\ref{fig:spec}(bottom) shows compensated energy spectra. The low frequency part of the spectrum was compensated by $f^{1.7}$ (solid line), the middle range -- by $f^{2.8}$  (dashed) and the high frequency part -- by $\exp (a(f/f_0)^{0.5})$ (dashed-dotted). The combined compensated spectrum is indeed very flat up to $f_{\lambda_e}$.

\begin{figure}
\includegraphics[width=6.5cm]{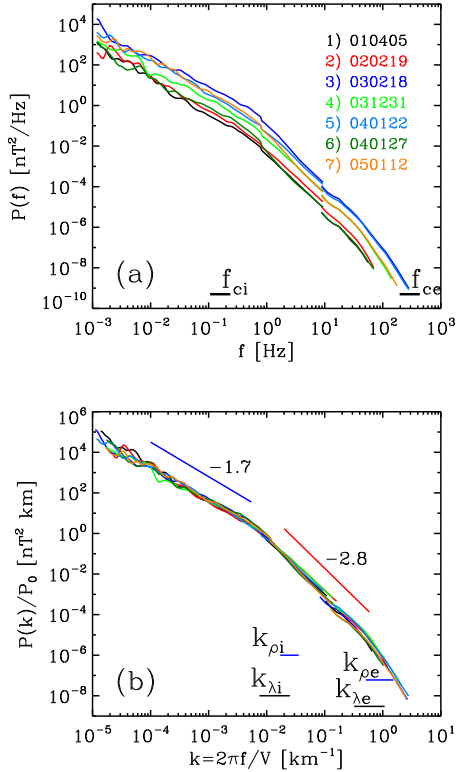}
\caption{\label{fig:spectra_freq}(a) Magnetic spectra for 7 time periods of 42 minutes;  spread of $f_{ci,e}$ for the 7 intervals is shown (b) $k$-spectra normalized over $P_0$;  characteristic wave numbers, $k_{\rho_i}=1/\rho_i$ etc., are shown. 
}
\end{figure}

The spectra for the seven intervals are presented in Figure~\ref{fig:spectra_freq}(a).  Horizontal bars indicate the spread of $f_{ci,e}$ among these seven independent observations. One can see that the spectra have similar shapes. Their intensity is, however, different. To superpose the spectra, we begin by applying Taylor's hypothesis, which should be valid for the whole frequency range, as far as quasi-parallel whistler waves are not observed during selected intervals (as  mentioned above). Thus, we assume that the frequency--spectra are indeed Doppler shifted $k$--spectra $P(k)=P(f)V/2\pi$ with $k=2\pi f/V$.  Then, we determine a relative intensity of the $j$-th spectrum, $S_j$ with $j=1,..., 7$, as  $P_{0}(j)=\langle S_j/S_1\rangle$, where $S_1$ is a reference spectrum, and $\langle\cdot \rangle$ indicates a mean over the range of wave vectors $10^{-5}<k<10^{-1}$~km$^{-1}$.  With this normalization, the  rescaled spectra may be nearly superposed   as shown in  Figure~\ref{fig:spectra_freq}(b).   
 
\begin{figure}
\includegraphics[width=8.0cm]{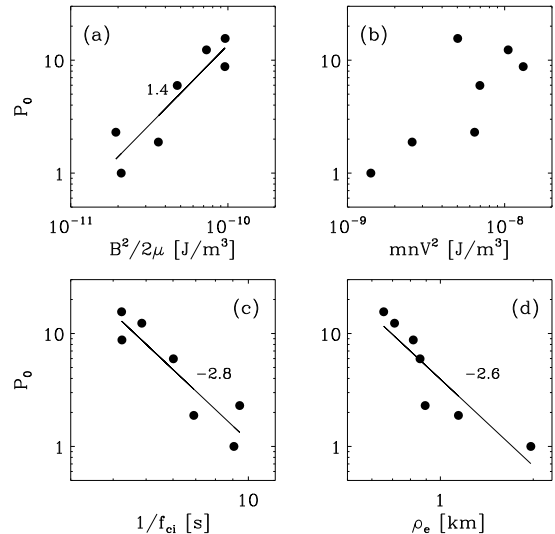}
\caption{\label{fig:P0_rhoe}  
Relative spectral intensity $P_0$ as a function of (a) magnetic and (b) kinetic energies;  (c) $P_0$ as a function of the ion cyclotron period and  (d) the electron gyroradius. Linear fits with corresponding slopes are shown by solid lines. }
\end{figure}

One expects that the spectral level, $P_0$, depends on  the solar wind kinetic, thermal  or magnetic energy. The scatter plots shown in  Figure~\ref{fig:P0_rhoe}(a) and (b) indicate clearly a power law dependance of $P_0$ on the magnetic energy, and a  less clear dependance on the kinetic energy, and  thermal energy (not shown). 

To understand  the meaning of the observed dependence on the magnetic energy, one may use a Kolmogorov-like phenomenology. Suppose first that the solar wind magnetic turbulence dissipates through an effective diffusion mechanism of  $\sim \eta \Delta B$ ($\eta$ being a -- probably turbulent -- magnetic diffusivity), and second, that the observed turbulence is quasi stationary.  In such a case, there is a balance between the energy input from nonlinear interactions  at large scales and the energy drain from the dissipation at small scales. This implies that  the energy transfer rate $\epsilon$ depends on the dissipation scale $\ell_d$ as  $\epsilon = \eta^3 \ell_d^{-4}$; thus $P_0 \sim \epsilon^{2/3} \sim \ell_{d}^{-8/3}$.  The dependences observed in Figure~\ref{fig:P0_rhoe}(c) and (d), $P_0\sim (1/f_{ci})^{-2.8}$ and $P_0 \sim \rho_e^{-2.6}$,  are  very close to the prediction of this phenomenological model. More statistics are needed to confirm the observed exponents. We can state,  however, that the observed dependences imply that $\rho_e$ and/or $f_{ci}$ and/or $f_{ce}$ play an important role in the dissipation processes in collisionless plasmas.   
Let us now confirm these results.

\begin{figure}
\includegraphics[width=8.5cm]{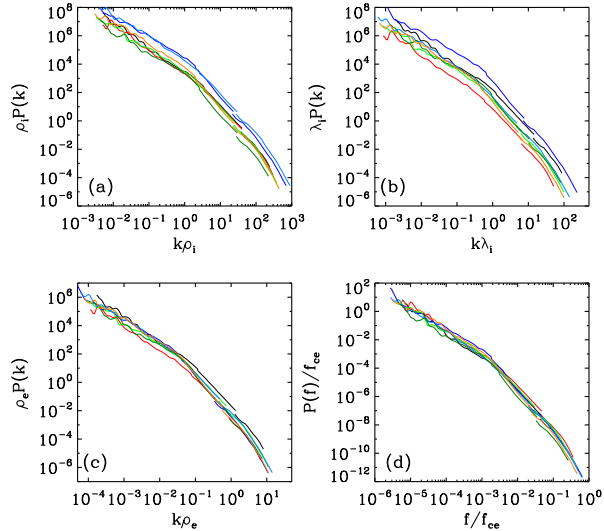}
\caption{\label{fig:spec_x_norm}  
Universal Kolmogorov function $\propto \ell_d E(k)$ for hypothesized
dissipation scales $\ell_d$ as a function of (a) $k\rho_i$, (b) $k\lambda_i$, (c) $k\rho_e$ and (d) $f/f_{ce}$. } 
\end{figure}

From the balance between the energy input and the dissipation, for the Kolmogorov's spectrum $E(k)$, it follows as well that  $E(k)\ell_d/\eta^2$  is a universal function of $k\ell_d$ \cite{Frisch1995,Davidson04book}. Figure~\ref{fig:spec_x_norm} tests which of the kinetic scales is to be used as $\ell_d$ to recover a universal function from the observed spectra. We assume for simplicity that $\eta$ is constant, despite the varying plasma conditions. One can see that the $\rho_i$ and $\lambda_i$--normalizations are not efficient to collapse the spectra together. Normalization on $\lambda_e$ gives the same result as for $\lambda_i$.  At the same time, the normalization on $\rho_e$ and $f_{ce}$ bring the spectra close to each other. This confirms that the electron gyroradius $\rho_e$ and/or cyclotron periods of the particles are important in the dissipation. 

With the present observations it is not possible to distinguish between $\rho_e$ and cyclotron periods as far as there is a clear correlation between $\rho_e$ and $B$. We can argue, however,  that if the cyclotron period had been the only dissipation scale, the turbulent cascade  would have stopped by the cyclotron damping of Alfv\'en waves at $f_{ci}$ showing an exponential cut-off at this scale \cite{Stawicki2001}. Solar wind  observations  show the contrary:  the turbulent spectrum continues up to 
electron scales. Thus, we conclude that  $\rho_e$ is the dissipation scale of magnetic turbulence in the solar wind, but we can not exclude that at $f_{ci}$ and $f_{ce}$ there is a partial dissipation by 
cyclotron damping.  

In the present letter we analyzed  high resolution magnetic spectra from MHD to electron scales.  We show here for the first time that whatever the  plasma conditions and  the solar wind regime, slow or fast, the magnetic spectra have similar shape. This indicates a certain universality, at least for the quasi-perpendicular configuration between ${\bf B}$ and ${\bf V}$.    Such a quasi-universal spectrum consists of three parts: two power-laws and an exponential domain. At MHD scales it follows a Kolmogorov's $\sim k^{-5/3}$ spectrum, in agreement with previous observations. Between $f_{ci}$ and Doppler shifted $\lambda_i$ and $\rho_i$ a spectral break is observed. Above the break, it follows a $k^{-2.8}$ power-law. At scales, of the order of $k\rho_e\simeq (0.1-1)$ it follows an exponential $\exp(-a(f/f_0)^{1/2})$. This is the first observation of an exponential magnetic spectrum in space plasmas. Such spectra were predicted by the anisotropic dissipation model  of \citet{Gogoberidze2005}. The author suggests that small scale  fluctuations with oblique ${\bf k}$ are diffused on oblique fluctuations from the inertial range. This diffusion is anisotropic and it gives an $\sim \exp(-k^{\Delta \alpha/2})$ spectrum in the dissipation range, where $\Delta\alpha = \alpha_{\perp}-\alpha_{\|} $ is the difference between the energy diffusion scaling perpendicular and parallel to ${\bf B}$. 

It is a long standing problem to distinguish between the role of different kinetic  scales in space plasmas. We  show for the first time that the role of dissipation scale in space plasma turbulence is played by the electron gyroradius, as assumed by several previous authors \cite{Leamon1999, Schekochihin2009,Fouad09}.  

\begin{acknowledgments}
We wish to acknowledge discussions with R.~Grappin, P.~Demoulin, N.~Meyer and S.~Bale. 
\end{acknowledgments}

\bibliography{aps-biblio-sw-short}
\end{document}